\documentclass{article}

\usepackage{blindtext}
\usepackage{authblk}
\usepackage{titlesec}
\date{}

\usepackage[english]{babel}

\usepackage[letterpaper,top=2cm,bottom=2cm,left=3cm,right=3cm,marginparwidth=1.75cm]{geometry}

\usepackage{amsmath}
\usepackage{graphicx}
\usepackage[colorlinks=true, allcolors=blue]{hyperref}

\usepackage[font=small,labelfont=bf,
   justification=justified,
   format=plain]{caption}

\usepackage{setspace}
\usepackage[english]{babel}

\usepackage[letterpaper,top=2cm,bottom=2cm,left=3cm,right=3cm,marginparwidth=1.75cm]{geometry}

\usepackage{amsmath}
\usepackage{graphicx}
\usepackage[colorlinks=true, allcolors=blue]{hyperref}

\usepackage[font=small,labelfont=bf,
   justification=justified,
   format=plain]{caption}

\title{Suppressing Plasmonic Heating in Aqueous Environments with Hexagonal Boron Nitride}

\author[1,2]{Martina Russo}
\author[1]{Roland van der Vegt}
\author[1]{Bohai Liu}
\author[1]{Sam Beijers}
\author[1]{Sara Salera}
\author[3]{Guillaume Baffou}
\author[1,4,*]{Klaas-Jan Tielrooij}
\author[1,2,**]{Peter Zijlstra}
\affil[1]{Eindhoven University of Technology, De Groene Loper 19, Eindhoven 5612 AZ, The Netherlands}
\affil[2]{Institute of Complex Molecular Systems, Eindhoven University of Technology, 5600 MB Eindhoven, The Netherlands}
\affil[3]{Institut Fresnel, CNRS, Aix Marseille Univ, Centrale Med, Marseille, France}
\affil[4]{Catalan Institute of Nanoscience and Nanotechnology (ICN2), CSIC and BIST, Campus UAB, Bellaterra (Barcelona), 08193, Spain}
\affil[*]{Email: k.j.tielrooij@tue.nl}
\affil[**]{Email: p.zijlstra@tue.nl}

\begin{document}
\maketitle

\newpage

\begin{abstract}

Optical heating of plasmonic nanostructures, caused by light absorption and following non-radiative decay, is a critical challenge in nanoscale systems. Although plasmonic effects enable enhanced optical functionalities, the associated temperature rise can degrade performance in heat-sensitive applications such as biosensing, nanophotonics, and microelectronics. Conventional cooling strategies fail at these scales due to limited heat transport and high interfacial thermal resistance, motivating the integration of advanced materials for thermal management. Here, we investigate hexagonal boron nitride (hBN) thin flakes as nanoscale heat spreaders to mitigate plasmonic heating of gold nanospheres when they are immobilized on hBN deposited on glass and surrounded by water. Using finite-element simulations, we quantify the influence of hBN thickness, in-plane thermal conductivity, and interfacial thermal conductance on cooling efficiency. Complementary experiments employ cross-grating wavefront microscopy (CGM) for nanothermometry to map temperature distributions around optically heated gold nanoparticles and quantify the cooling effect of hBN flakes. To study the effect of hBN thickness, we extend the application of CGM for rapid, non-invasive, and all-optical thickness characterization of non-absorbing 2D materials. Our results reveal a strong thickness dependence, where heat dissipation in thin flakes is limited by the heat capacity of hBN and in thick flakes by interfacial thermal conductance. We obtain a reduction in the steady-state temperature rise by up to $\sim$60 \% compared to glass by including hBN. In addition, the presence of two main heat dissipation pathways emerges: a direct one from the nanoparticle to the hBN and an indirect one from the particle via water to the hBN. This combined simulation–experiment framework offers a versatile approach to improve thermal management in plasmonic systems and beyond, establishing design guidelines for integrating 2D materials into thermally sensitive platforms such as biosensors, integrated circuits, and energy conversion.

\end{abstract}

\section*{Keywords}

Plasmonic heating, hexagonal boron nitride, gold nanoparticles, nanoscale thermal management, interfacial thermal conductance, nanothermometry, heat spreading

\newpage

\section*{Introduction}

Plasmonic nanostructures have attracted considerable attention due to their ability to confine and enhance electromagnetic fields at the nanoscale through localized surface plasmon resonances (LSPRs). When metallic nanoparticles, particularly gold and silver nanostructures, are illuminated at their plasmon resonance frequency, the collective oscillation of conduction electrons leads to strong optical absorption and scattering. A significant fraction of the absorbed optical energy is dissipated through non-radiative decay pathways involving Landau damping, electron–electron scattering, and subsequent electron–phonon coupling, which ultimately converts optical energy into heat within picoseconds \cite{Boriskina2017}. This process results in localized temperature increases around the nanoparticle, commonly referred to as plasmonic heating \cite{Baffou2020}. While this phenomenon has been exploited \cite{Govorov2007} in applications such as photothermal therapy \cite{Baffou2020}, catalysis \cite{Baldi2023}, and solar energy harvesting, it can also introduce detrimental thermal effects in many nanoscale photonic systems. In particular, excessive local heating can induce thermal drift, modify refractive indices, degrade molecular probes, or even damage nearby structures, thereby limiting the performance of plasmon-enhanced biosensors, nanophotonic devices, and integrated microelectronic platforms \cite{Pop2006}. These issues become especially critical in densely integrated architectures, where heat dissipation pathways are restricted, and thermal accumulation can degrade device stability and signal fidelity. Understanding and controlling heat generation and dissipation in plasmonic nanostructures has therefore emerged as a central challenge in thermoplasmonics and nanoscale photonics \cite{Jones2018}.

Over the past decade, several strategies have been proposed to quantify and mitigate plasmonic heating. Early studies focused on modeling heat generation and diffusion around individual nanoparticles, highlighting the strong dependence of the temperature rise on particle size, absorption cross-section, and the thermal properties of the surrounding medium. Under continuous-wave (CW) illumination, steady-state heating typically results from the balance between optical absorption and thermal diffusion into the environment, whereas pulsed excitation produces transient non-equilibrium dynamics in which ultrafast electron heating precedes lattice thermalization \cite{Baffou2011}. The thermal dynamics of plasmonic nanoparticles under pulsed excitation can involve sub-picosecond electron–electron scattering followed by electron–phonon coupling and phonon relaxation, processes that govern the ultimate temperature rise and heat transfer to the surrounding medium. Although numerous theoretical and experimental efforts have attempted to quantify nanoscale temperature distributions \cite{Baffou2011}, accurately measuring and controlling local temperature fields remains challenging due to the sub-diffraction-limited spatial scales involved and the complex interplay between local and collective heating effects. Consequently, despite significant advances in thermoplasmonics, effective strategies for controlling heat generation and dissipation at the nanoscale remain limited, particularly in realistic environments such as aqueous media or integrated photonic platforms.

To address these limitations, increasing attention has been devoted to integrating plasmonic nanostructures with materials exhibiting high thermal conductivity to facilitate heat spreading and improve thermal management. Various substrates and heat-spreading layers, including crystalline oxides, ceramics, and carbon-based materials, have been explored to mitigate heat accumulation in plasmonic nanostructures by facilitating lateral heat diffusion and improving thermal coupling with the environment \cite{Hoque2021}. Oxide substrates such as sapphire \cite{Setoura2013} \cite{Panais2023} or magnesium oxide offer optical transparency and moderate thermal conductivity, making them compatible with photonic devices; however, their thermal conductivities are typically one to two orders of magnitude lower (34 W m$^{-1}$ K$^{-1}$ for sapphire \cite{Rahman2019} and 45-60 W m$^{-1}$ K$^{-1}$ for magnesium oxide \cite{Slack1962}) than those of the most efficient heat spreaders, which limits their ability to dissipate localized nanoscale hotspots generated under intense optical excitation \cite{Cahill2003} \cite{Slack1973}. High-conductivity ceramics such as diamond exhibit exceptional thermal transport properties and have therefore been proposed as efficient thermal platforms for nanoscale sensing and photonic devices \cite{Tanos2020}. Nevertheless, diamond substrates often require complex fabrication processes, are costly, and are difficult to integrate with heterogeneous nanoscale architectures or solution-processed nanostructures. Carbon-based materials, particularly graphene and related nanostructures, provide extremely high in-plane thermal conductivities and have attracted considerable interest as ultrathin heat spreaders \cite{Sang2019} \cite{Balandin2008} \cite{Balandin2011}. Despite these advantages, graphene presents several drawbacks in plasmonic and optoelectronic systems, including non-negligible optical absorption, electrical conductivity that may interfere with device operation, and strong electronic coupling with metallic nanostructures that can modify their optical response. In addition, heat dissipation in graphene-based systems is often limited by large thermal boundary resistance at interfaces with substrates or surrounding media \cite{Koh2010}. These limitations highlight the need for alternative materials that combine high thermal conductivity with optical transparency, chemical stability, and electrical insulation. 

Hexagonal boron nitride (hBN) is particularly promising in the context of 2D materials for thermal management \cite{Wu2022} \cite{Alborzi2021}. Owing to its wide band-gap of $\sim$6 $\mathrm{eV}$, hBN is essentially transparent in the visible range while maintaining a high in-plane thermal conductivity of 250 to 650 W m$^{-1}$ K$^{-1}$ \cite{Wang2011} \cite{Tang2024} \cite{Yuan2019},  comparable to that of some carbon-based materials \cite{Lindsay2011}. Furthermore, its atomically layered structure enables the fabrication of ultrathin flakes that can be integrated with plasmonic nanostructures without introducing optical losses or electronic interactions \cite{Song2010}. These properties make hBN an attractive platform for nanoscale thermal management and motivate systematic investigations of its heat-spreading capabilities in nanoscale systems \cite{Bohai2026}. Experimental and theoretical investigations have shown that hBN can support efficient phonon-mediated heat transport and significantly influence thermal conduction across nanoscale interfaces \cite{Bohai2026} \cite{Gargiulo2023} \cite{Caldwell2019} \cite{Sichel1976}. Recent work by Liu et al. \cite{Bohai2026}, focusing on both transient heating dynamics and steady state temperature increase, has demonstrated that hBN encapsulation reduced the temperature ramp rate of electrically biased gold nanostrips by up to $\sim$40 \%. Nevertheless, despite its favorable thermal properties, the potential of hBN to control plasmonic heating remains unclear. In particular, how the thickness of hBN flakes, their intrinsic thermal conductivity, and the interfacial thermal conductance with both nanoparticles \cite{Cavigli2020} and the surrounding medium collectively determine heat dissipation from optically excited plasmonic nanoparticles remains elusive. Addressing these questions is essential for establishing design principles for integrating 2D materials as thermal management layers in plasmonic and nanophotonic devices.

In this work, we address these gaps through both simulations and experiments. Using finite-element modeling, we systematically investigate the effect of hBN thickness, in-plane thermal conductivity, and interfacial thermal resistance on the cooling efficiency of plasmonically heated gold nanospheres. To measure the temperature distributions of the gold nanoparticles, we employed cross-grating wavefront microscopy (CGM). CGM is a high-resolution, high-sensitivity, wavefront imaging technique that enables measuring the optical path difference (OPD) induced by the refractive index change with sub-nanometer sensitivity, and translates it into the corresponding temperature increase \cite{Baffou2012} \cite{Baffou2023}. In addition, we demonstrate the application of CGM for rapid, non-invasive thickness characterization of hBN flakes, generalizing its application to non-absorbing 2D materials. 

Our results demonstrate up to $\sim$60 \% reduction in the equilibrium temperature rise and a strong dependence of the cooling efficiency on hBN thickness, in-plane thermal conductivity, and interfacial thermal conductance. We identify two regimes in the thickness dependence, with heat dissipation for thin flakes being limited by the hBN heat capacity, and for thick flakes being limited by the interfacial conductance. Our study also reveals insights into the fundamental mechanism of heat dissipation, with a direct pathway dominant at the nanoparticle-hBN interface, and an indirect pathway from the nanoparticle to the surrounding water and then into the hBN. By integrating simulations and experiments, we establish design guidelines for the use of hBN as a nanoscale heat spreader and demonstrate its potential for efficient thermal management that is directly applicable to biosensing, microelectronics, nanophotonics, and energy conversion platforms. 

\section*{Results and Discussion} 

\subsection*{hBN sample preparation for cooling of gold nanoparticles}

\begin{figure} [h!]
\centering
\includegraphics[width=0.7\linewidth]{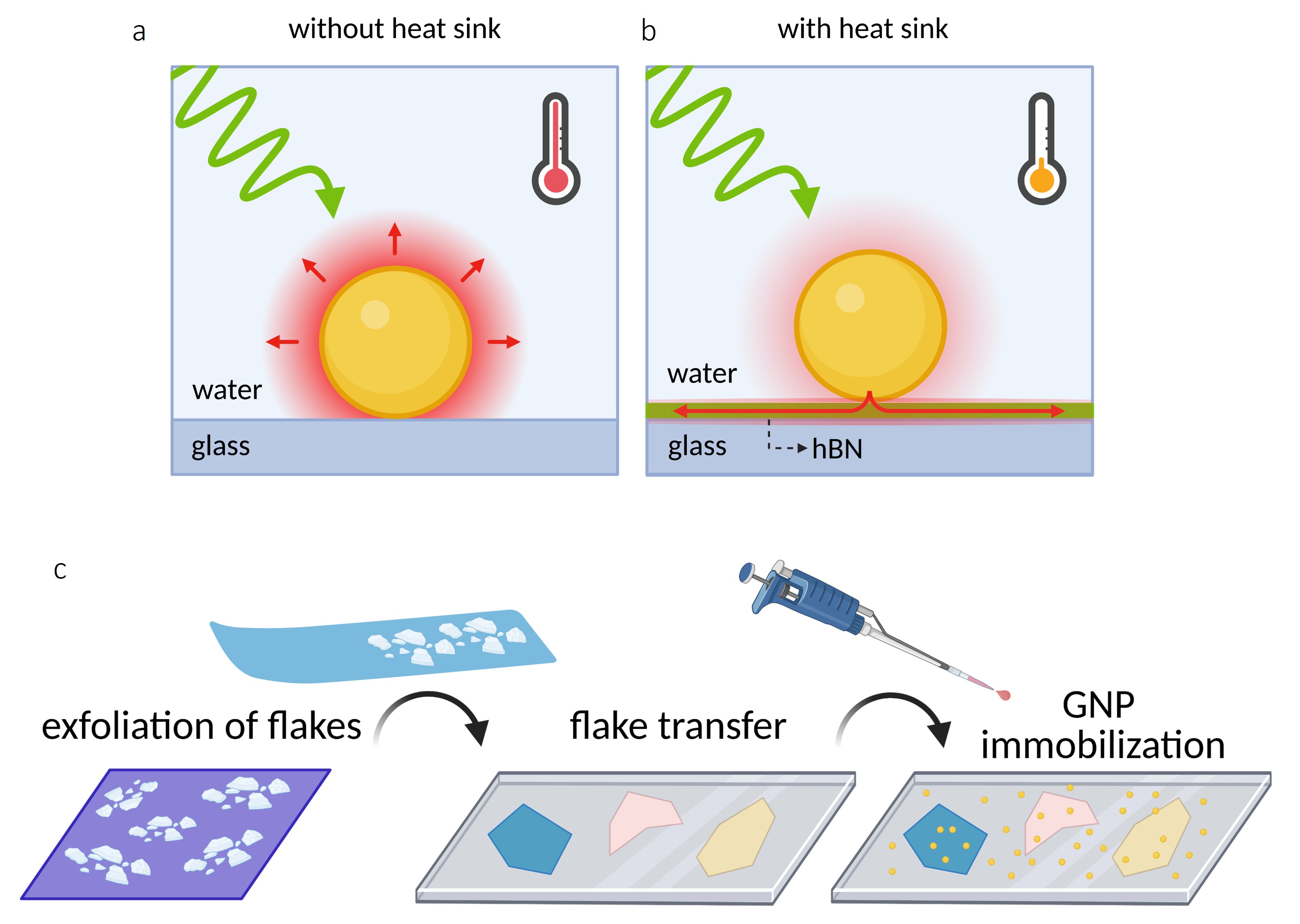}
\captionof{figure}{Illustration of the concept of laser heating and thermal management by hBN, and of the sample preparation. Plasmonic nanoparticles on glass (a) irradiated by a laser undergo an intense temperature increase. The addition of an hBN underneath the particles (b) results in an efficient heat dissipation, leading to a lower temperature increase. (c) Cartoon illustrating the sample preparation workflow. hBN flakes of different thicknesses (different colors in the cartoon) are exfoliated from hBN powder onto SiO$_2$ substrates, and subsequently transferred onto glass coverslip using a dry-transfer procedure. A solution of GNPs is then drop-casted onto the glass coverslip and rinsed after the incubation time. The resulting samples contain nanoparticles that are immobilized both on the hBN flakes and on the bare glass coverslip.}
\label{Fig1}
\end{figure}

Figure \ref{Fig1} illustrates the core concept of our proposed thermal management strategy and the experimental approach. Figure \ref{Fig1}a shows the reference configuration, where a gold nanoparticle immobilized on glass and surrounded by water undergoes significant optical heating upon laser irradiation. The heat accumulates locally due to the low thermal conductivity of glass and water ($k_\mathrm{SiO_2}=1.38$ W m$^{-1}$ K $^{-1}$ \cite{Yang2024} and $k_\mathrm{H_2O}=0.6$ W m$^{-1}$ K$^{-1}$ \cite{Martin1933}), resulting in a pronounced temperature rise. In contrast, Figure \ref{Fig1}b depicts the modified configuration with an hBN flake inserted between the nanoparticle and the glass substrate. The high in-plane thermal conductivity of hBN enables rapid lateral heat spreading, followed by conduction into the substrate, thereby reducing local temperature increases.

The samples employed in this research consisted of 100 nm GNPs immobilized either on a bare glass slide or on hBN flakes deposited on a glass slide (detailed hBN sample preparation and GNPs immobilization procedures in Sections 1 and 2 of Supporting Information). The workflow of the sample preparation (summarized in Figure \ref{Fig1}c) started with the mechanical exfoliation of hBN flakes from hBN powder onto a freshly cleaned highly doped Si wafer with a thermally grown oxide layer of 285 nm \cite{Blake2007}, using adhesive tape.

 We then transferred the flakes onto a clean glass coverslip using a polydimethylsiloxane (PDMS) cube covered with a thin polycarbonate (PC) film, as in reference \cite{Pizzocchero2016}. After cleaning the surface to remove any residual PC or contamination, we drop-casted a solution of 100 nm diameter gold nanospheres and incubated it for 20 minutes, resulting in efficient particle immobilization both on the hBN flakes and on the bare glass surface. After rinsing, we mounted a fluidic chamber on the sample and filled it with water, and we placed a second glass slide on top of it to seal the cell and prevent water evaporation during the measurements. 

\subsection*{Numerical simulations}

\begin{figure} [h!]
\centering
\includegraphics[width=1\linewidth]{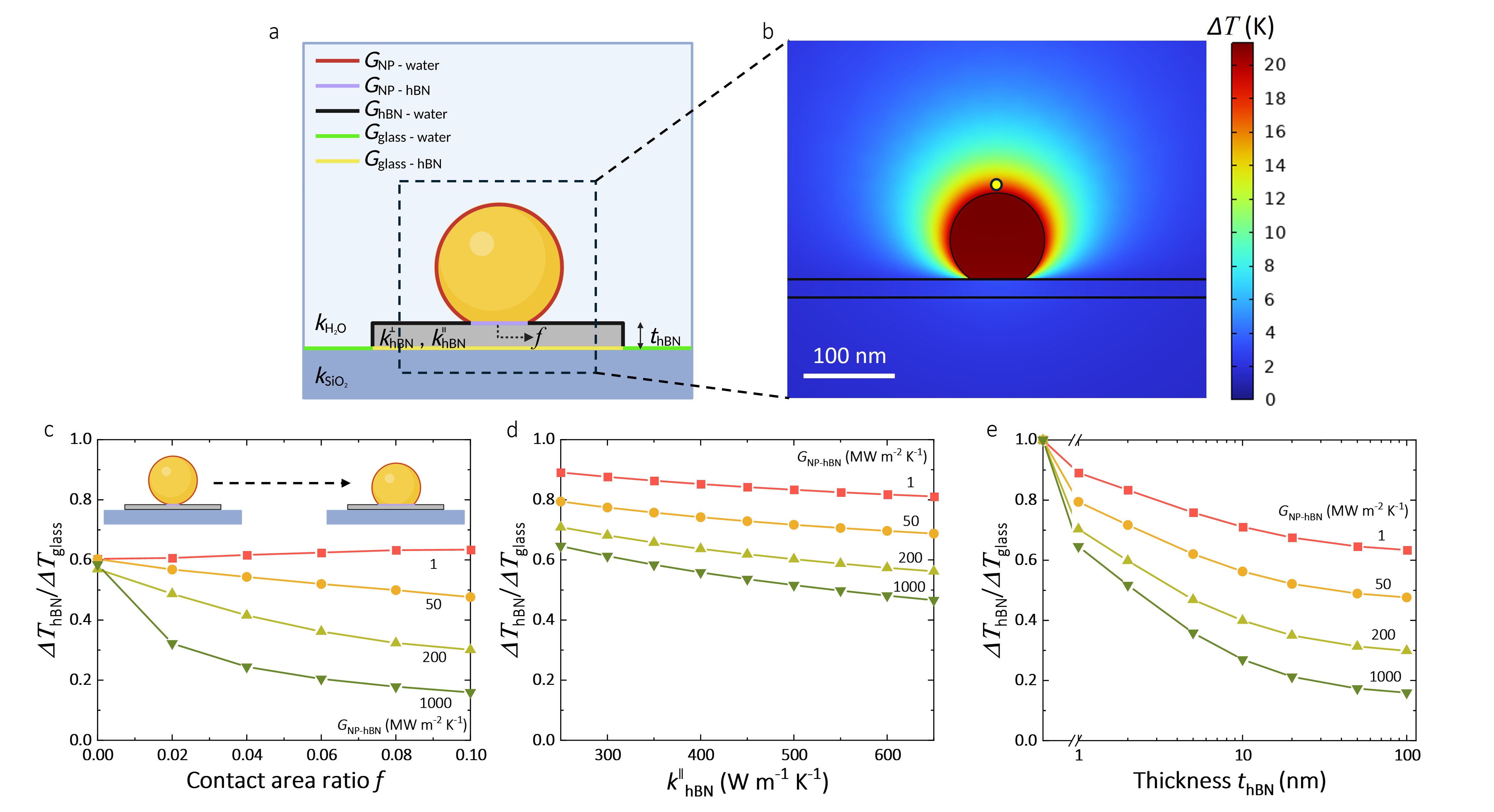}
\captionof{figure}{Simulations framework and main results. (a) Illustration of the simulated system, consisting of a 100 nm diameter GNP immobilized on a hBN flake, with thickness $t_\mathrm{hBN}$ and in-plane conductivity $k^{\parallel}_\mathrm{hBN}$, on glass. The contact area between the GNP and the hBN is described by the fraction $f$. The figure depicts the interfacial thermal conductance $G$ of all the interfaces in the system. (b) Example distribution of the temperature increase for $t_\mathrm{hBN}=20$ nm. We probed the temperature increase at a 5 nm distance above the nanosphere. (c) Cooling factor, ($\frac{\Delta T_\mathrm{hBN}}{\Delta T_{SiO_2}}$), as a function of the contact area ratio $f$ and for varying interfacial resistance $G_\mathrm{NP-hBN}$ (legend in (d)) ($t_\mathrm{hBN}=100$ nm, $k^{\parallel}_\mathrm{hBN}=250$ W m$^{-1}$ K$^{-1}$). (d) Cooling factor as a function of the in-plane hBN conductivity $k^{\parallel}_\mathrm{hBN}$ ($t_\mathrm{hBN}=1$ nm, $f=0.1$). (e) Cooling factor as a function of the hBN thickness $t_\mathrm{hBN}$ ($k^{\parallel}_\mathrm{hBN}$ = 250  W m$^{-1}$ K$^{-1}$, $f=0.1$).}
\label{Fig3}
\end{figure}

To investigate the effect of hBN on the thermal profiles surrounding optically heated gold nanoparticles and quantify its cooling efficiency, we performed numerical simulations using Finite Element Modeling (FEM) in COMSOL under steady-state heat transfer in solids. We modeled a 100 nm diameter GNP immobilized on a glass substrate or on an hBN flake of thickness $t_\mathrm{hBN}$, with the whole system immersed in water. The latter system includes five interfaces with interfacial thermal conductance \cite{Lyeo2006} $G$: gold–water, gold–glass (or gold–hBN), water–glass, water–hBN \cite{Alosious2022}, and hBN–glass, as schematized in Figure \ref{Fig3}a. 

To mimic realistic and finite particle–substrate contact, we treated the nanoparticles as truncated spherical particles, and we defined the contact area fraction $f$ as the ratio between the gold–substrate and gold–water surface areas. Although colloidal gold nanoparticles are frequently described as nanospheres, structural studies have demonstrated that, due to the face-centered cubic (fcc) crystal structure of gold and the anisotropy of surface energies, nanoparticles minimize their total surface energy by forming faceted polyhedral shapes \cite{Marks1994}. As confirmed by high-resolution transmission electron microscopy \cite{Kuo2004}, gold nanoparticles typically expose low-index facets such as \{111\} and \{100\}, resulting in geometries described as polyhedra composed of nanoscale facets \cite{Xia2009} (quantification of average facet length from TEM images in Section 7 of Supporting Information).  

We treated the hBN flake as thermally anisotropic, with in‑plane conductivity $k^{\parallel}_\mathrm{hBN}$ and cross‑plane conductivity $k^{\perp}_\mathrm{hBN}$. The in-plane and cross-plane conductivities are set respectively to 250 W m$^{-1}$ K$^{-1}$ and 8 W m$^{-1}$ K$^{-1}$ where not otherwise specified \cite{Jaffe2023}. We modeled glass and water as isotropic media, with $k_\mathrm{SiO_2}=1.38$ W m$^{-1}$ K $^{-1}$ \cite{Yang2024} and $k_\mathrm{H_2O}=0.6$ W m$^{-1}$ K$^{-1}$ \cite{Martin1933}.

To enable direct comparison across parameter sweeps, we first defined the temperature probe location at 5 nm above the nanoparticle apex in water. We then defined the cooling factor as the temperature rise at the probe location in the presence of the hBN layer $\Delta T_\mathrm{hBN}$ normalized to the temperature rise on glass without heat spreader $\Delta T_\mathrm{SiO_2}$, meaning the ratio $\frac{\Delta T_\mathrm{hBN}}{\Delta T_\mathrm{SiO_2}}$. Figure \ref{Fig3}b shows a representative steady-state temperature map for a 100 nm GNP on hBN. 

The simulated temperature map here shows that, for contact area and hBN thickness sufficiently large, as quantified next, the heat is mainly conducted through the gold-hBN interface and rapidly dissipated along the in-plane direction. As a result, the heat is confined to the nanoparticle and its surrounding liquid environment, while the heat sink is capable of dissipating the heat flux without experiencing any temperature increase.

We first varied $f$ while sweeping the thermal boundary conductance $G_\mathrm{NP-hBN}$ over three orders of magnitude (1–1000 MW m$^{-2}$ K$^{-1}$), and Figure \ref{Fig3}c summarizes the resulting cooling factor. Even for $f=0$, meaning for a point-contact between the particle and the hBN layer, the cooling factor is already significant, with a $40\ \%$ decrease of the absolute temperature increase with respect to glass. 

We interpret this finding by identifying two main heat transfer pathways as underlying thermal physics mechanisms. An indirect one is the heat transfer from the nanoparticle to the surrounding water, and from there to the hBN heat sink, which turns out to be the dominant one in the specific case of $f=0$. Despite the limited contact area and no direct heat dissipation through the gold-hBN interface, the hBN layer efficiently absorbs the heat through the water-hBN interface and efficiently dissipates it laterally, resulting in a relevant cooling factor. The other pathway is the direct heat conduction through the gold-hBN interface, followed by lateral dissipation from the hBN film. This pathway is dominant for larger contact areas, for which the cooling factor further decreases.  

Investigating the effect of the gold-hBN interfacial conductance, two regimes emerge. The first one is an interface-limited regime, with low $G_\mathrm{NP-hBN}$. Here, the increase in $f$ has little impact because the heat flux across the interface is minimal, limited by the low interfacial conductance. This results in a poor heat transfer, and thus a high absolute temperature reached by the particle. The second regime is an interface-favored one, with high $G_\mathrm{NP-hBN}$. Here, the heat flux across the nanoparticle-hBN layer interface becomes dominant, translating into an effective heat transfer from the particle to the heat sink. Due to the high in-plane thermal conductivity, hBN efficiently spreads the heat laterally, reducing the $\frac{\Delta T_\mathrm{hBN}}{\Delta T_\mathrm{SiO_2}}$ ratio further as the contact area increases.
These results emphasize that both a sufficient interfacial conductance and contact area are required to fully exploit the thermal heat sink.

Next, we quantified how the cooling factor depends on $k^{\parallel}_\mathrm{hBN}$ (Figure \ref{Fig3}d). We explored a range of in-plane conductivities from 250 to 650 W m$^{-1}$ K$^{-1}$, in agreement with what has been reported for supported and free-standing hBN flakes \cite{Wang2011} \cite{Tang2024} \cite{Yuan2019}. Simulations show that, as expected, an increase in the in-plane conductivity leads to an enhanced cooling efficiency, with a lower absolute temperature increase due to the enhanced heat dissipation of hBN along the in-plane direction.

Finally, we varied $t_\mathrm{hBN}$ over a $1-100$ nm range at fixed $k^{\parallel}_\mathrm{hBN}$ and $f$, obtaining the results in Figure \ref{Fig3}e. We varied the thickness-dependent cross‑plane conductivity $k^{\perp}_\mathrm{hBN}$ for each simulated thickness, with the values taken from the study of Jaffe et al \cite{Jaffe2023} on the thickness-dependent cross-plane conductivity of hBN, ranging from $k^{\perp}_\mathrm{hBN}=0.6$ W$^{-1}$ K$^{-1}$ for $t_\mathrm{hBN}=1$ nm to $k^{\perp}_\mathrm{hBN}=6.9$ W$^{-1}$ K$^{-1}$ for $t_\mathrm{hBN}=100$ nm.

The results show that even a few-layer flake with a thickness of 1 nm yields a measurable cooling effect due to lateral spreading, but the effect increases with thickness, as larger thickness translates into a larger cross-section available for in‑plane conduction, or equivalently a larger heat capacity and thus more heat dissipation. For very thin flakes, in fact, the insufficient thickness partially limits the benefit of the enhanced $k^{\parallel}_\mathrm{hBN}$; on the other hand, tens of nanometers provide an efficient cooling effect (20-70 $\%$), with $k^{\parallel}_\mathrm{hBN}$ values (250 W m$^{-1}$ K$^{-1}$) approaching the performance of much thicker isotropic heat spreaders \cite{Setoura2013}. 

Overall, the simulations indicate that hBN can achieve significant cooling factors when sufficient thickness and heat capacity enable efficient in-plane heat transport, and the contact area is optimized to allow efficient heat conduction from the heated nanoparticle into the heat sink. This motivates the experimental approach presented in the following section, where we used CGM nanothermometry and flake thickness characterization to quantitatively validate the simulation results.

\subsection*{CGM for nanothermometry and hBN characterization}

\begin{figure} [h!]
\centering
\includegraphics[width=1\linewidth]{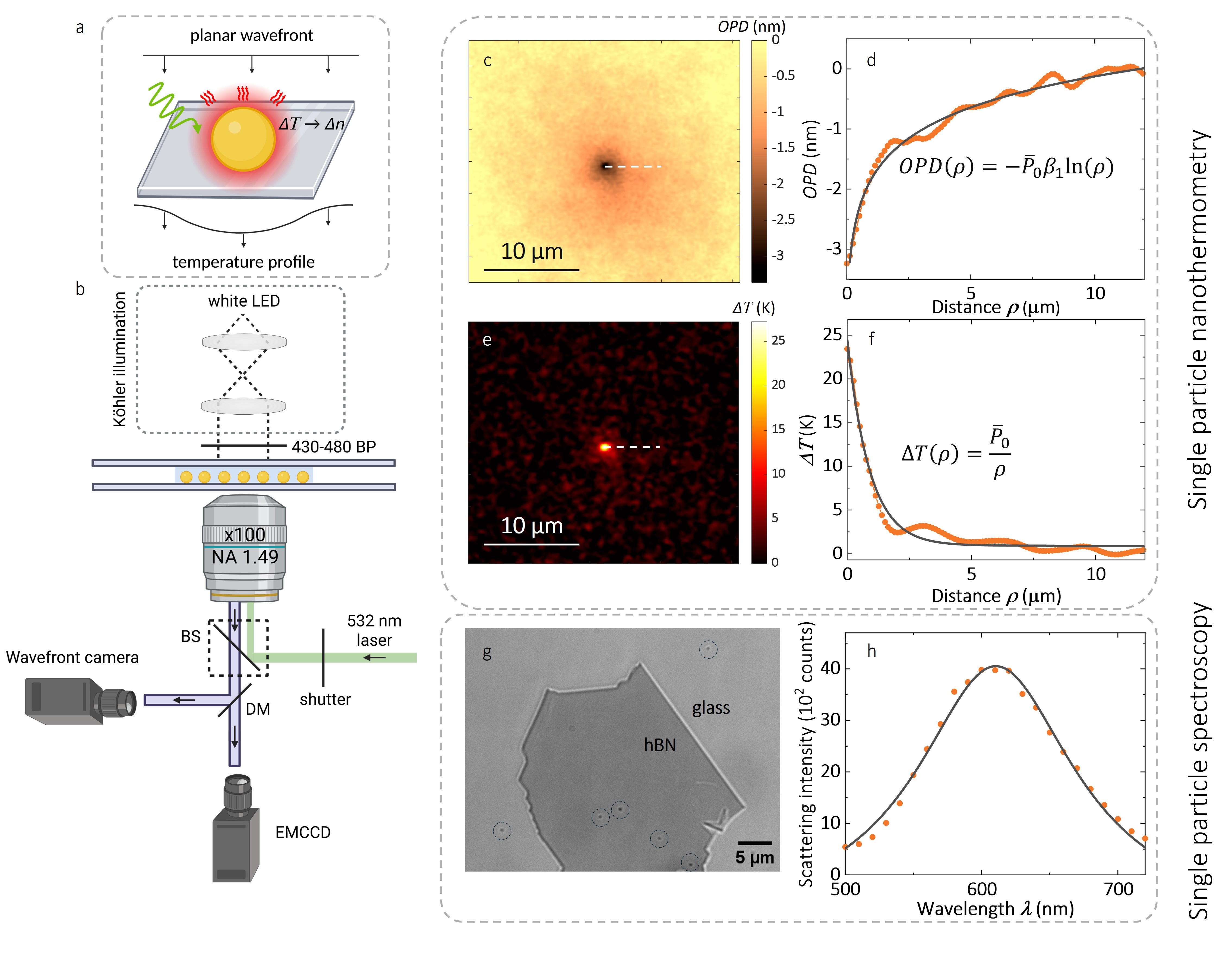}
\captionof{figure}{Experimental setup and measurements workflow. (a) Schematic working principle of the CGM technique and its application to measure the temperature increase of optically heated plasmonic nanoparticles. (b) Schematic of the optical setup used for CGM measurements. (c) Raw image of the OPD of a 100 nm GNP irradiated by the heating laser. (d) Measured radial profile (orange data points) of the OPD distribution in image c (dashed line) as a function of the radial distance from the center of the particle, and the theoretical description using the equation in the figure (solid black line). (e) Temperature increase distribution resulting from the OPD in image c. (f) Radial profile (orange data points) of the temperature distribution in image e (dashed line) as a function of the radial distance from the center of the particle, and the theoretical description using the equation in the figure (solid black line). (g) Transmission image of a sample consisting of a 100 nm thick hBN flake on glass, with GNPs immobilized both on the flake and glass marked by circles. (h) Example of experimental scattering spectra of the employed GNPs on hBN (orange data points) and superimposed Lorentzian description (solid black line).}
\label{Fig2}
\end{figure}

The cartoon in Figure \ref{Fig2}a summarizes the working principle of quadriwave lateral shearing interferometry (QLSI), the technique employed for both nanothermometry and hBN thickness characterization. QLSI is a high-resolution, high-sensitivity, wavefront imaging technique. When implemented in a microscope, in a configuration named cross-grating wavefront microscopy (CGM), it can yield label-free quantitative characterization of transparent objects lying in the field of view of a microscope, such as biological cells in culture \cite{Bon2009}, nanomaterials \cite{Baffou2023}. In particular, it can quantify the optical properties of 2D materials, such as the complex refractive index, with a single-shot acquisition. CGM can also be used as a temperature microscopy technique, since microscale temperature gradients usually lead to refractive index gradients responsible for wavefront distortions \cite{Baffou2012}. 

We used both these modalities in this study, in parallel. In addition to these key strengths and to the temperature measurement accuracy below 1 K, CGM, as a diffraction-limited optical technique, has the advantage of retrieving the average temperature map in a small volume around the nanoparticle. Compared to other thermometry techniques for single particles heated with incident light \cite{Martinez2024}, such as anti-Stokes or Raman thermometry, this results in an ideal probing region. A direct measurement of the inner temperature of the particle would not allow access to the temperature measurements in its surroundings and, as a consequence, to the interpretation of the mechanism behind the heat transmission and dissipation in the system, for which a probing point outside the particle surface is instead required and provided by CGM.

Figure \ref{Fig2}b depicts the CGM optical setup used for the nanothermometry and flake thickness characterization measurements. We used a white light-emitting diode (LED) in a K\"ohler configuration to illuminate the sample from the top. We inserted a bandpass filter into the optical path to set the probe bandwidth of the phase-sensitive camera. The light was transmitted through a 100$\times$ magnification, 1.49 NA objective, and split by a dichroic mirror to allow simultaneous visualization on an electron multiplying charge-coupled device (EMCCD camera) and the nanothermometry measurements. 
We used a 532 nm excitation laser to heat the GNPs, by coupling it from the bottom of the sample through a dichroic mirror (details on the laser beam characterization in Section 3 of Supporting Information). We employed the EMCCD camera both to monitor the one-photon photoluminescence (1-PL) of the GNPs in real-time during the nanothermometry measurements (Section 4 of Supporting Information) and to perform single-particle hyper-spectral scattering microscopy (HSM) measurements (Section 5 of Supporting Information). We used these two methods to identify single particles and monitor their stability during the experiments. 

We performed the nanothermometry measurements by synchronizing the CGM camera with a beam shutter (Thorlabs SH05R/M) placed in the excitation path using a MATLAB script. The synchronization allows for acquiring a reference interferogram when the shutter is closed, and particles are not irradiated by the heating laser, and a measurement interferogram when the shutter is open, and particles irradiated by the laser are heated up. For every single measurement, we recorded 100 reference frames and 100 measurement frames and averaged them to yield one final averaged reference and measurement interferogram. The exposure time of the camera was set to 100 ms, resulting in a total duration of less than 30 s per measurement acquisition. This procedure ensures the optimization of the noise level of the measurements, with a value of $\sigma_\mathrm{OPD}\ \sim$1 nm comparable to literature results \cite{Chaumet2024} \cite{Marthy2022}, and corrects for artifacts potentially caused, for example, by mechanical drift of the sample or fluctuation of the light sources.

 Figure \ref{Fig2}c shows the obtained raw OPD data from the CGM measurements of an irradiated GNP. The resulting OPD profile (Figure \ref{Fig2}c-d) across the single particle agrees with the expected theoretical curve that describes the spatial distribution of the optical path difference due to thermal-induced variation of the refractive index distribution around a point-like source of heat. Figure \ref{Fig2}e-f show the respective 2D map and line profile for the resulting temperature increase of the same particle. Because heating changes the refractive index via the thermo-optic coefficient $\beta_1=dn/dT$, using the known $\beta_1$ of the medium ($\beta_1=-1.0514\ \times 10^{-4}$) and a heat-diffusion model, assuming a steady-state temperature profile around the heat source, we inverted the measured OPD map to obtain the spatial temperature distribution \cite{Baffou2021}. 

The data are again in agreement with the predicted temperature profile as a function of the radial distance from the center of the particle. For the temperature increase data analysis in the following, it is relevant to note that we measured multiple temperature radial profiles for a single particle, extracting the temperature increase profile along multiple lines centered on the particle and rotating one with respect to the other. The resulting rotational average was then taken into account by averaging all the individual profiles. We then fitted the averaged profile with a Gaussian function, and the background-corrected amplitude was used to extract the temperature increase corresponding to the measurement (Section 6 of Supporting Information). 

Figure  \ref{Fig2}g shows the transmission bright field image of one field of view (FOV), where both particles on glass and on hBN are visible. We used the same optical setup, with some modifications in the excitation path (Section 5 of Supporting Information), to measure the scattering spectra of the immobilized particles. The characteristic spectrum of a 100 nm GNP is shown in Figure \ref{Fig2}h, where the experimental data-points are described using a Lorentzian function. The results are in agreement with the theoretical expectation of the scattering spectrum for a single gold nanosphere of the corresponding size, ruling out the presence of clusters or other impurities among the measured objects on the sample surface, whose heating and temperature profile are not described by a point-like source of heat model. 

\subsection*{hBN thickness characterization}

Accurate determination of the hBN flake thickness is essential for correlating the thermal performance with material properties. We implemented CGM as a rapid, non-invasive way of measuring the hBN flake thickness, following the principles described by Khadir et al. \cite{Khadir2017}. For layered materials such as hBN and other 2D materials, the OPD is directly related to the flake thickness through its effective refractive index. 

\begin{figure} [h!]
\centering
 \includegraphics[width=1\linewidth]{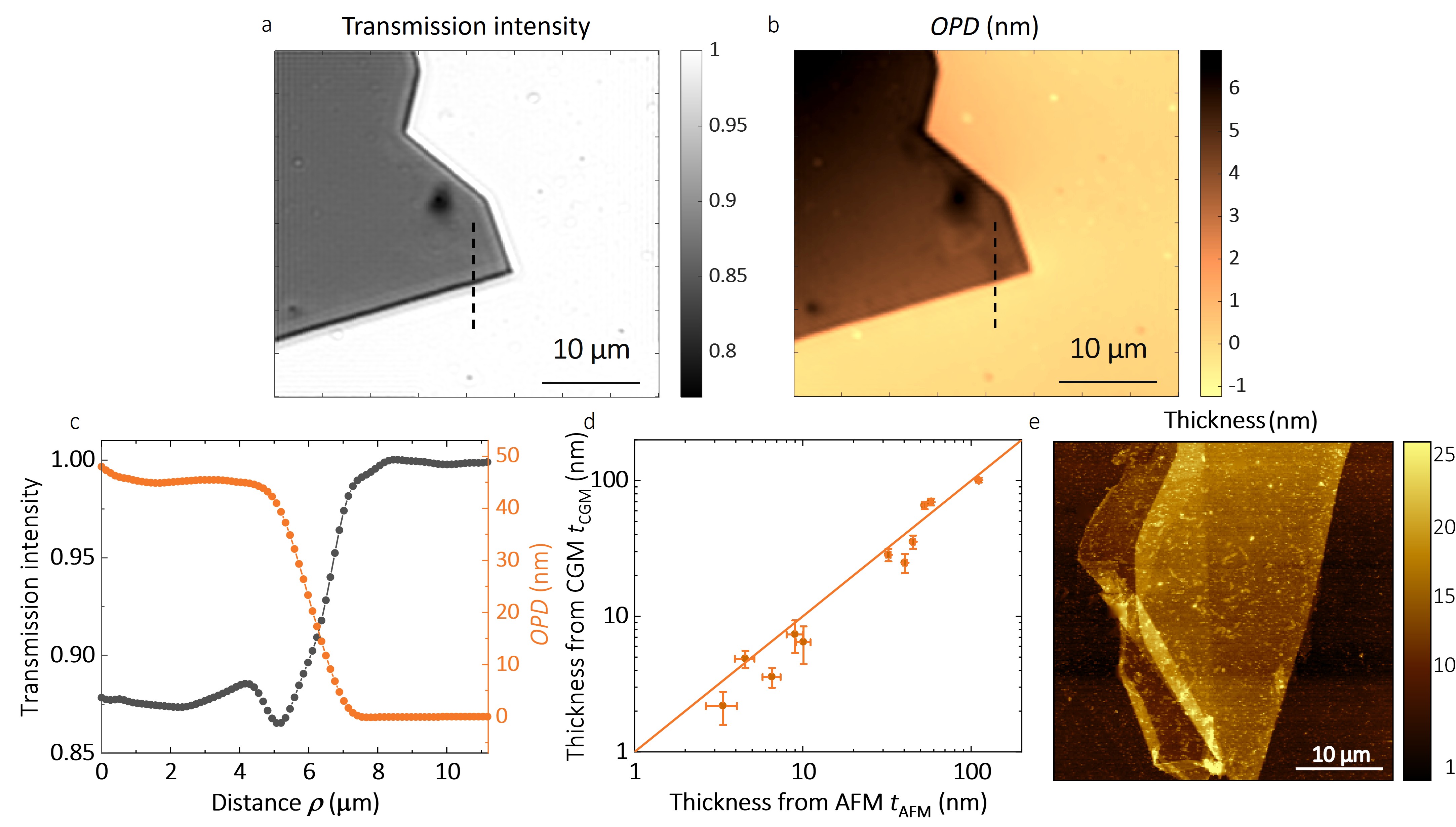}
\captionof{figure}{Workflow of the CGM data processing to characterize the hBN flakes thickness. (a) Transmission intensity (a) and OPD map (b) of a hBN flake ($t_\mathrm{hBN}=37$ nm). (c) Profile of the transmission intensity and OPD along the dashed lines in figures (a) and (b). (d) Comparison of the thickness characterization obtained from CGM and AFM measurements. Each data point represents one of the 11 measured flakes, with thickness ranging from 3 to 110 nm (error bars represent 68 $\%$ confidence interval). (e) AFM measurement of a hBN flake of graph (d) ($t_\mathrm{hBN}=8.9$ nm).}
\label{Fig4}
\end{figure}

This enables a rapid, all-optical and non-invasive thickness characterization without the need for mechanical probing in AFM measurements \cite{Nemes2008}, time-consuming Raman spectroscopy \cite{Jin2020} or optical contrast methods requiring multi-parameter fits with poorly constrained parameters \cite{Golla2013}. 

Another advantage of the CGM technique in this case is the possibility of using regular transparent substrates, without the need for absorbing substrates such as SiO$_2$/Si wafer, commonly employed to measure the optical reflectivity of hBN flakes \cite{Golla2013}.  

Figure \ref{Fig4} illustrates the complete workflow. Figure \ref{Fig4}a shows the transmission intensity image of an hBN flake on a glass substrate, while Figure \ref{Fig4}b displays the corresponding OPD map obtained from CGM. Both maps indicate a clear contrast between the flake and the substrate, with OPD values increasing with flake thickness. The profiles of transmission intensity and OPD along the dashed lines in Figure \ref{Fig4}a and \ref{Fig4}b are plotted in Figure \ref{Fig4}c, highlighting the sharp transition at the flake edge.

We extracted the hBN thickness by fitting the OPD values to a transfer-matrix model that accounts for the refractive index of hBN \cite{Grudinin2023} and the underlying substrate, as described in Khadir et al \cite{Khadir2017}. This approach enables direct conversion of OPD into physical thickness (mathematical model and equations in Section 8 of Supporting Information). Figure \ref{Fig4}d compares the thickness values obtained from CGM with those measured by AFM for a set of flakes spanning a $3-110$ nm thickness range. Every data point in the graph represents a single flake, with the resulting thickness from the two methods and the error bar, both on the AFM and CGM characterization. The strong correlation, close to the identity line, confirms the reliability of the GCM-based method. Figure \ref{Fig4}e shows an example AFM image of one of the measured flakes, for reference.

The CGM-based approach demonstrated an accuracy within 3 nm compared to AFM, while significantly reducing acquisition time and avoiding sample contact. This capability is particularly advantageous for studies requiring high-throughput characterization or integration with optical setups, as in our thermometry experiments. By combining phase-based thickness determination with transmission contrast, CGM provides a versatile tool for characterizing 2D materials in situ.

\subsection*{hBN cooling efficiency dependence on thickness}

\begin{figure} [h]
\centering
\includegraphics[width=1\linewidth]{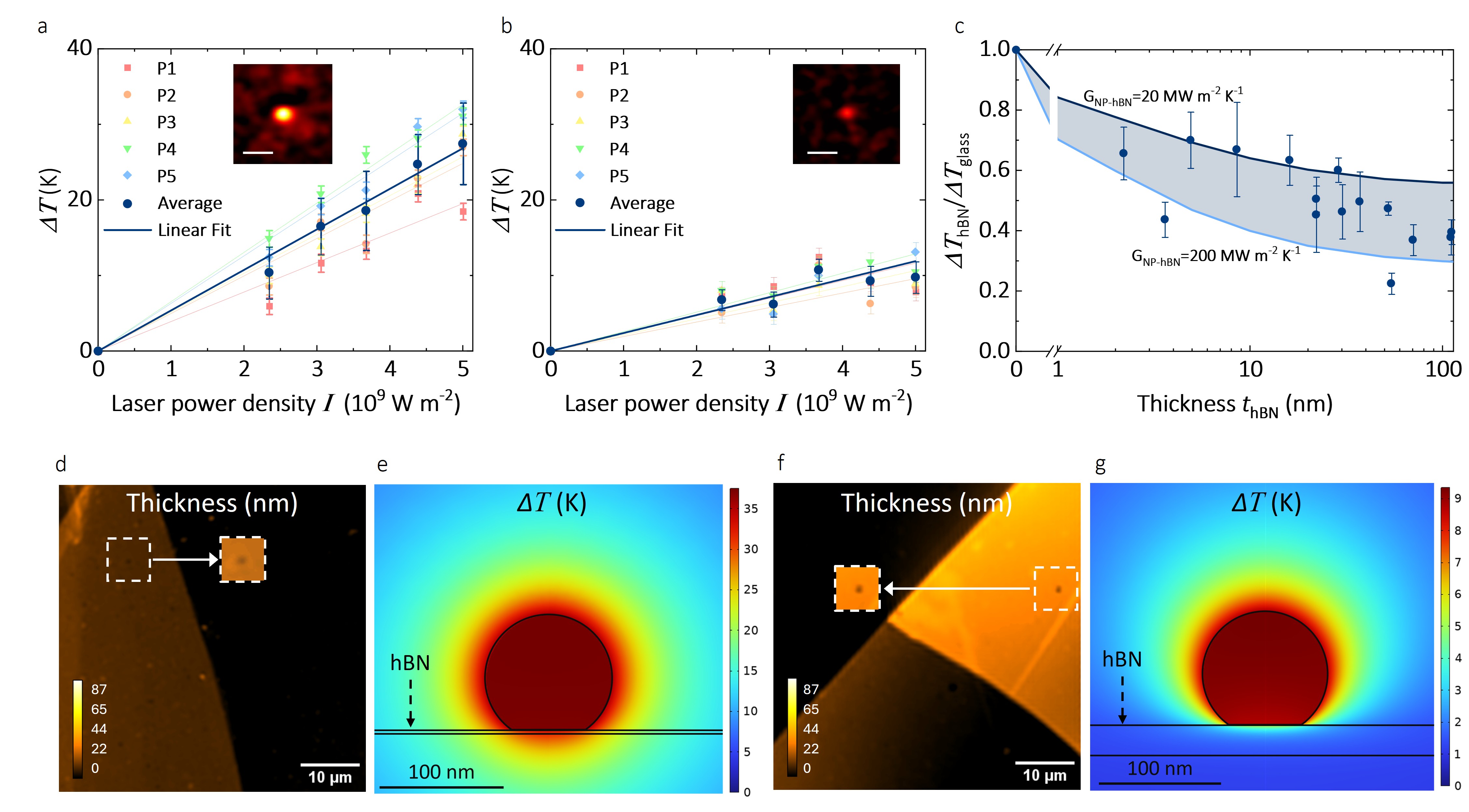}
\captionof{figure}{Overview of thickness dependence experimental results. Measured temperature increase of GNPs on glass (a) and hBN ($t_\mathrm{hBN}=52$ nm) (b) for increasing power densities. The average temperature increase of all the particles is described using a linear function. Insets show some example temperature maps for glass and hBN, respectively (scale bar of 1 $\mu m$). (c) Resulting experimental cooling factor as a function of the hBN flakes thickness from CGM. The data are compared with the simulations shown in Figure 3e, considering the standard deviation of the data-points (68 \% confidence interval within the lines). (d) OPD map of a thin flake ($t_\mathrm{hBN}=8.9$ nm), with visible low phase contrast compared to the glass substrate. Zoom in on a particle, with a rescaled calibration bar for clarity. (e) Simulated temperature profile on a $t_\mathrm{hBN}=1$ nm hBN flake ($G_\mathrm{NP-hBN}=200$ MW m$^{-2}$ K$^{-1}$ , $f=1$, $k^{\parallel}_\mathrm{hBN}=250$ W m$^{-1}$ K$^{-1}$), showing a significant temperature increase of the substrate underneath the GNP. (f) OPD map of a thick flake ($t_\mathrm{hBN}=86$ nm), with high phase contrast compared to the glass substrate. Zoom in on a particle on the flake. (g) Simulated temperature profile on a $t_\mathrm{hBN}=25$ nm hBN flake ($G_\mathrm{NP-hBN}=200$ MW m$^{-2}$ K$^{-1}$, $f=1$, $k^{\parallel}_\mathrm{hBN}=250$ W m$^{-1}$ K$^{-1}$), showing the limited temperature increase of the thick hBN heat sink, with low temperature increase confined to the GNP and its surroundings in the liquid environment.}
\label{Fig5}
\end{figure}

To experimentally assess the cooling effect provided by hBN flakes, we performed nanothermometry measurements using CGM on GNPs immobilized on glass and on hBN flakes of varying thickness. Figure \ref{Fig5} shows one example of the measured temperature increase $\Delta T$ as a function of laser power density for GNPs on glass (Figure \ref{Fig5}a) and on hBN (Figure \ref{Fig5}b), respectively. Each dataset includes measurements from five individual particles, with the average temperature increase described by a linear function, consistent with the expected relation between absorbed power and steady-state temperature rise. Insets display representative temperature maps, highlighting the pronounced thermal profile arising around the GNP on glass compared to the reduced heating observed for the same particle on hBN.

We quantified the cooling effect of hBN, defined above, as a function of the hBN thickness, ranging from a few nanometers to around 100 nm. The experimental data in Figure \ref{Fig5}c reveal a clear trend: even ultrathin hBN layers provide measurable cooling, while thicker flakes significantly enhance heat dissipation, reducing the steady-state temperature rise by up to $\sim$60 \% compared to glass. These results align closely with the finite-element simulations presented in Figure \ref{Fig3}e, which predict a monotonic improvement in cooling with increasing thickness. 

The comparison between experimental points and simulation curves (Figure \ref{Fig3}e) for two representative interfacial conductance values, $G_\mathrm{NP-hBN}=20$ MW m$^{-2}$ K$^{-1}$ and $G_\mathrm{NP-hBN}=200$ MW m$^{-2}$ K$^{-1}$, with a fixed contact area with $f=0.1$, highlights the critical role of interfacial properties in achieving optimal heat transfer to the heat sink substrate, together with a thickness high enough for maximum thermal management performance. 

To elucidate the cooling mechanism, we compared the experimental data with simulation results, zooming in on the GNP-hBN interface region. A thin hBN flake, visible with a low phase contrast in the OPD map in Figure \ref{Fig5}d, results in a limited cooling efficiency, as visible from the relevant temperature increase of both particle and substrates in Figure \ref{Fig5}e. As a consequence, the absolute steady state temperature of the nanoparticle and its liquid surrounding is significant, up to 35 K above ambient, and the absolute temperature increase of the hBN and glass substrate underneath the GNP is comparable to that of the particle itself. The limited heat capacity of the thin hBN layer is notably responsible for the heat conduction from the particle into the glass substrate, with both the heat flux pathways, the direct one from the nanoparticle to the heat sink and the indirect from the nanoparticle to the water and then to the heat sink, being hampered. On the other hand, a thick hBN flake, visible with high phase contrast in Figure \ref{Fig5}f, results in significant cooling of the GNP. An example temperature map for a simulation with a thick flake is shown in Figure \ref{Fig5}g. Here, the increased heat sink thickness provides an enhanced heat capacity to efficiently dissipate heat from the nanoparticle, reducing the resulting maximum temperature increase of the particle, the heat sink itself, and the underlying glass. In this case, in fact, the hBN layer confines the heat profile inside it and does not propagate it to the glass substrate. 

Overall, these findings confirm that hBN acts as an effective nanoscale heat spreader, due to its high in-plane thermal transport that efficiently redistributes heat laterally. The agreement between simulations and experiments validates the design guidelines established earlier and demonstrates the feasibility of integrating hBN for thermal management in plasmonic systems, with particular focus on hBN thickness, nanoparticle-hBN contact area, and interfacial conductance. In particular, this framework allows for a better understanding of the role of the gold nanoparticle-hBN interfacial conductance, whose quantification or direct measurement is currently challenging and poorly investigated due to experimental limitations \cite{Park2016} \cite{Swartz1989} \cite{Stoner1993} such as atomically thin interfaces, extremely small temperature drops at the interface, and imperfect interfaces due to fabrication defects or impurities. A systematic study and measurement of the GNP-hBN contact area, for example using other shapes of nanoparticles with known facets geometry and size, or inducing changes in the nanosphere contact area through controlled annealing experiments \cite{Kosutova2025} (Section 9 of Supporting Information), would aid in reducing the number of unknown parameters of the system, paving the way for an indirect quantification of the nanoparticle-heat sink interfacial conductance.

\section*{Conclusions}

In this work, we addressed the challenge of mitigating optical heating in plasmonic systems by integrating hexagonal boron nitride flakes as nanoscale heat spreaders. Through our simulation and experimental approach, we demonstrate that hBN significantly reduces the temperature rise of optically heated gold nanoparticles, due to its high in-plane thermal conductivity, which efficiently dissipates heat laterally. Finite-element simulations reveal that cooling efficiency strongly depends on flake thickness, interfacial thermal conductance, and contact area, with thicker flakes and optimized interfacial heat transfer achieving cooling factors comparable to those of much thicker isotropic heat spreaders. Our study also highlights the limiting factors of the thermal management provided by thin hBN films, with effective heat dissipation being restrained by the heat capacity of hBN in thin flakes, and by the interfacial thermal conductance in thick flakes. Complementary nanothermometry measurements using CGM validated these predictions, showing up to $\sim$60 \% reduction in temperature increase for hBN flakes of $\sim$100 nm thickness. In addition, our work represents an application of CGM for rapid, non-invasive thickness characterization of hBN flakes, generalizing its application to non-absorbing 2D materials, in addition to non-transparent ones. This dual functionality of CGM for both all-optical nanoscale nanothermometry and material characterization provides a powerful tool for high-throughput analysis of heat dissipation in 2D material-based architectures.

Altogether, our findings establish design guidelines for integrating hBN as an efficient thermal management layer in plasmonic systems and highlight its potential for broader applications in biosensing, nanophotonics, and microelectronics, where localized heating limits performance and reliability. This lays the basis for future work to investigate different particle shapes, to better control and improve the interfacial conductance, and to explore the use of different materials as a heat sink, to further enhance the cooling efficiency.

\section*{Experimental}

\subsection*{Sample preparation: hBN flakes exfoliation and transfer}

hBN flakes were prepared by mechanical exfoliation from hBN powder onto highly doped Si wafer with a thermally grown oxide layer of 285 nm, cleaned by oxygen plasma before the exfoliation. hBN powder was placed on adhesive tape and repeatedly folded and peeled to progressively thin the crystallites. The tape was then pressed onto the substrate and slowly removed, leaving exfoliated flakes on the surface.

A polymer stamp for dry transfer was fabricated by placing a polydimethylsiloxane (PDMS) cube on a rigid support substrate and covering it with a thin polycarbonate (PC) film prepared by spin coating. The PC film was picked up using adhesive tape with a circular opening and carefully positioned over the PDMS block to form the pickup interface.

Selected flakes were identified by reflection optical microscopy based on lateral size and optical contrast. Optical images were recorded to enable relocation during transfer and to verify flake integrity after processing. The transfer was carried out in a setup equipped with independently movable stamp and sample stages, the latter connected to a vacuum line for substrate fixation.

Flake pickup was performed at $110 \ ^\circ C$. The hBN/$SiO_2$ sample was aligned beneath the PC/PDMS stamp and raised until contact was established, while monitoring the advancing contact front. After maintaining full contact for 2 min, the stage was lowered, leaving the flake attached to the PC film. For deposition, the target substrate was heated to $180\ ^\circ C$ and brought into contact with the stamp following the same procedure. At this temperature, the PC film softens, enabling the release of the flake onto the substrate. Residual PC was removed by sequential washing in chloroform (5 min), incubation on a hotplate at $450\ ^\circ C$ (5 min, overnight washing in chloroform, and rinsing in isopropanol. Gold nanospheres were immobilized on the hBN flakes following the same procedure explained in the following section.

\subsection*{Sample preparation: GNPs immobilization on glass and hBN}

Borosilicate coverslips (thickness \#1.5) were sonicated in methanol for 20 min and dried by nitrogen flow. The coverslips were activated by plasma treatment for 1 min to hydrophilize the surface. 

A suspension of citrate-capped 100 nm diameter GNPs (A11-100-CIT-DIH-1-50, NanoPartz) was drop-casted on the glass substrates and incubated for 30 min, after which the coverslips were rinsed with Milli-Q water and dried under nitrogen flow.

An imaging spacer was placed on the coverslip, filled with 13 $\mu L$ of Milli-Q water, and sealed with a clean coverslip for the nanothermometry measurements.

\section*{Acknowledgments}

This project has received funding from the European Research Council (ERC) under the European Union’s Horizon 2020 Research and Innovation Program (Grant Agreement No. 864772). 

\section*{Supporting information}

Supporting information is available and includes: 

\begin{itemize}
 \item Experimental setup for hBN flakes exfoliation and transfer, heating beam characterization, 1-PL distribution, single-particle HSM of individual GNPs, temperature maps analysis, quantification of average GNPs facet length, mathematical model for hBN thickness characterization, annealing of GNPs.
\end{itemize}

\newpage

\begin{figure} [h!]
\centering
 \includegraphics[width=1\linewidth]{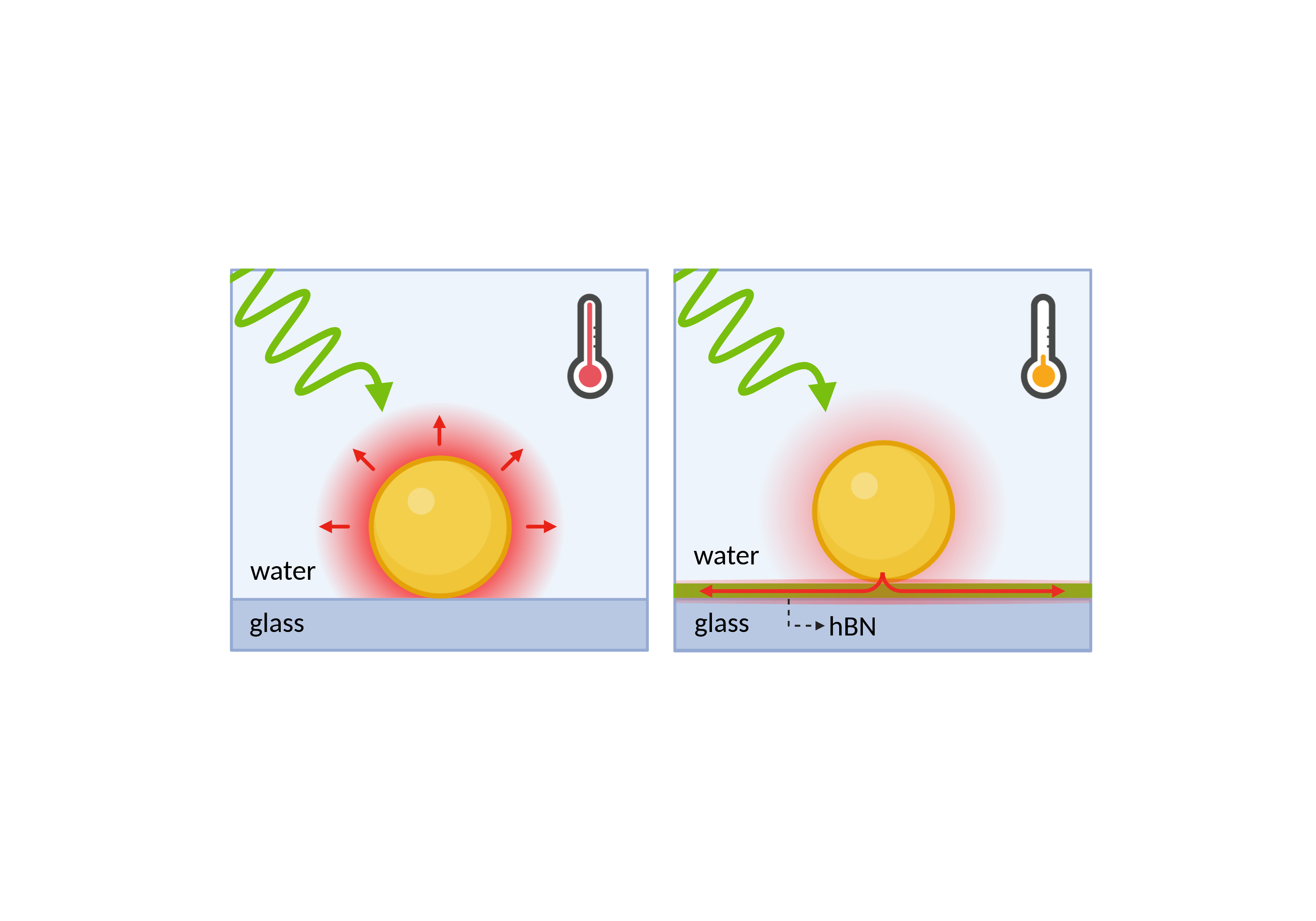}
\captionof{figure}{TOC graphic.}
\label{Fig4}
\end{figure}

\newpage  

\bibliographystyle{unsrt}
\bibliography{Bibliography}

\end{document}


\maketitle
\newpage

\tableofcontents

\newpage

\section{Experimental setup for hBN flakes exfoliation and transfer}

\begin{figure} [h!]
\centering
\includegraphics[width=1\linewidth]{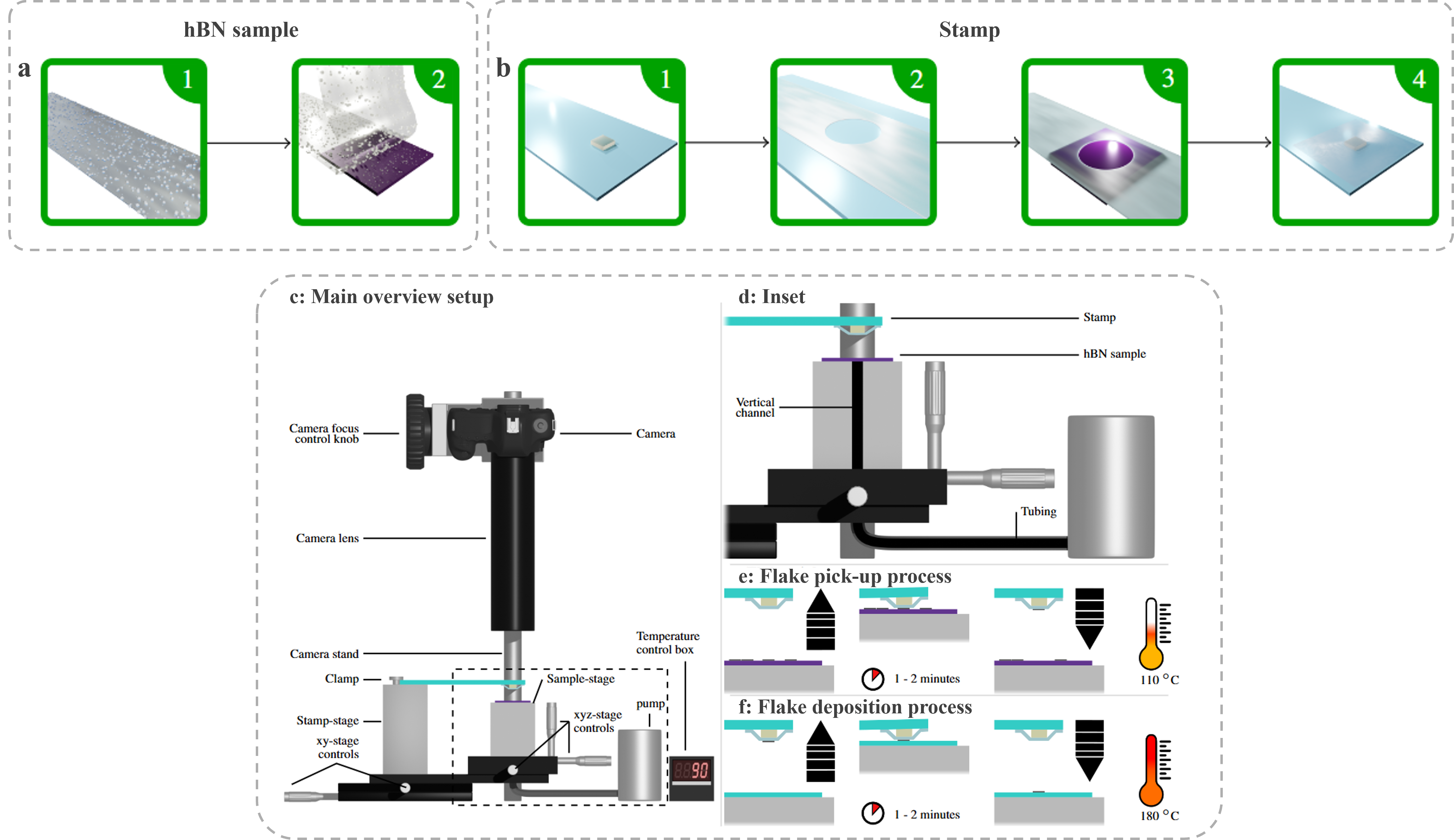}
\captionof{figure}{Overview of the steps for creating the hBN sample (a) and stamp (b) for dry transfer. (c) Schematic of the complete setup used for dry-transferring. (d) Enlarged
view of the stamp and sample in the setup; sliced sample-stage and tubing showcasing the airflow from the vertical sample-stage channel out through the pump. (e)) Schematic of the flake being transferred from the sample to the stamp. (f) Schematic of the flake being transferred from the stamp to the target-substrate.}
\label{S1}
\end{figure}

\section{Heating beam characterization}

To measure the heating beam size, one nanoparticle was moved across the beam, and the one-photon photoluminescence (1-PL) signal was recorded. The maximum 1-PL as a function of the displacement was plotted. The experimental data were fitted using a Gaussian function, and the FWHM obtained from the fit is taken as an estimate of the beam diameter.

\begin{figure} [h!]
\centering
\includegraphics[width=0.9\linewidth]{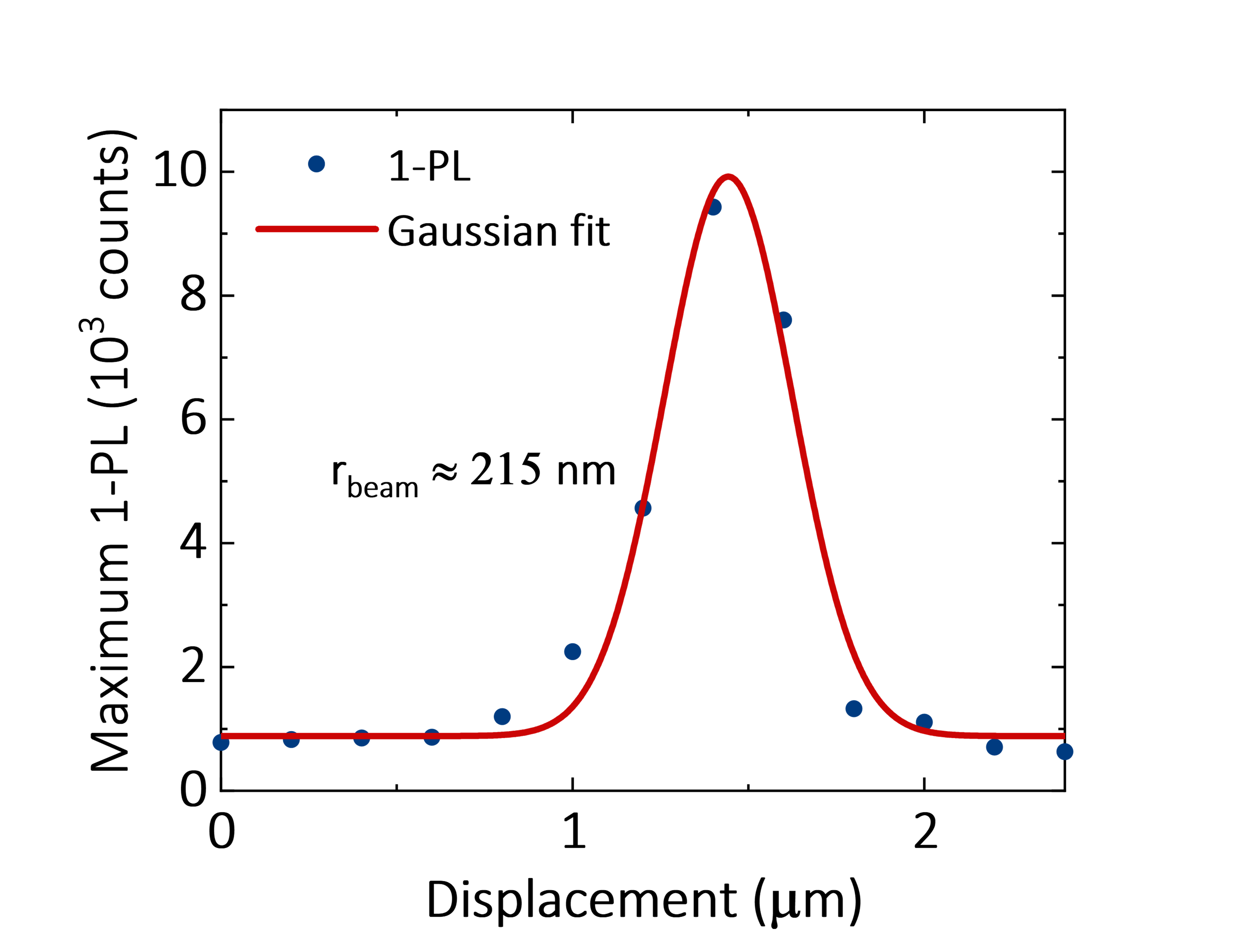}
\captionof{figure}{1-PL signal of one gold nanoparticle as a function of its position along a line cutting the center of the heating beam. The Gaussian fit of the data is superimposed, and the resulting beam radius is reported.}
\label{S2}
\end{figure}

\newpage

\section{1-PL distribution}

To ensure that any difference in the resulting temperature increase upon laser excitation for particles on glass and hBN does not come from a difference in the effective laser power density reaching the particles in the two cases, a study of the one-photon luminescence (1-PL) signal was conducted. A given number of particles on glass and hBN thick flake (100 nm) were irradiated by the heating laser with constant laser power, and the luminescence signal was collected by the EMCCD camera in the setup. The histograms of the resulting 1-PL signals intensity show no significant difference for the two substrates, confirming the even illumination condition and thus heating efficiency in the two cases.

\begin{figure} [h!]
\centering
\includegraphics[width=1\linewidth]{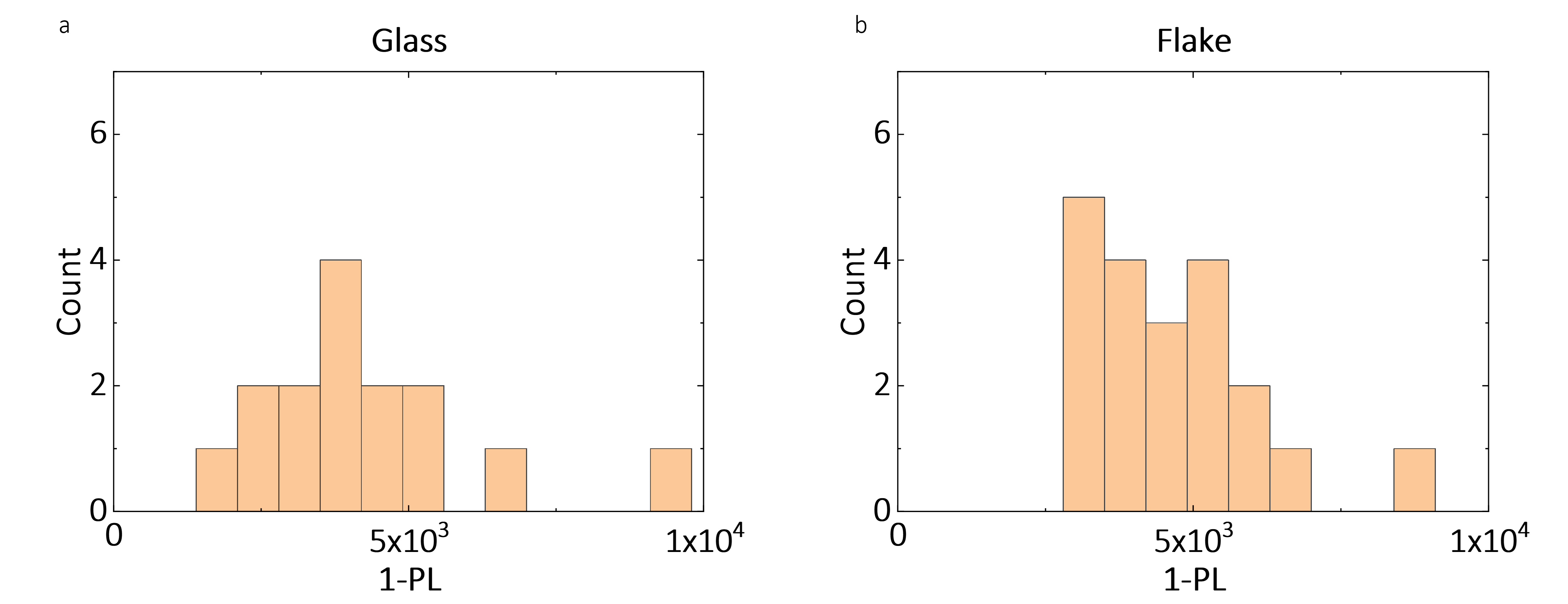}
\captionof{figure}{Histograms of the 1-PL signal intensities for GNPs on glass (a) and hBN thick flake (b) irradiated with the same laser power.}
\label{S3}
\end{figure}

\newpage

\section{Single-particle hyper-spectral scattering microscopy of individual GNPs}

We perform hyper-spectral scattering microscopy (HSM) to obtain scattering spectra of GNPs immobilized on glass and hBN flakes. The sample was illuminated by a fiber-coupled white-light source equipped with a tunable band-pass filter (SuperK COMPACT and SuperK VARIA tunable filter). Collimated by a Thorlabs RC08FC-P01, excitation light is sent to the sample with the same optical path used for fluorescence measurements.
The totally internal reflected beam is blocked by a custom-made beam-blocker located at the
bottom of the objective. Scattered light is collected in the detection path without filters
and focused. Excitation light was tuned from 500 nm to 720 nm with a bandwidth of 10 nm and set at 1$\%$ working power; for each step, the scattering intensity of the FoV was collected. Image
sequences were analyzed using custom Matlab software obtaining scattering spectrum of the region of interest in the FoV.

\section{Temperature maps analysis}

To extract the temperature increase of a single particle from each measurement, the temperature maps were processed as follows. The maximum temperature increase value of the map was extracted and taken as the center of 20 line profiles traced on the particle. The angular spacing between each line profile was constant to ensure a uniform coverage of the particle along each direction. The temperature profiles were extracted along each line and averaged together. The resulting average was then fitted with a Gaussian function with offset, and the amplitude minus offset was taken as the final temperature increase estimation. The Gaussian fitting function was used to describe the results in place of the $\sim$1/$\rho$ relationship to enable a more consistent and reproducible data analysis over different measurements. Moreover, the comparison between the results obtained with the two fitting functions was investigated. We observed a minimal discrepancy between the two methods, which turned out to be of the order of 1 K, and thus negligible and inside the error bar of the temperature measurement itself.

\begin{figure} [h!]
\centering
\includegraphics[width=1\linewidth]{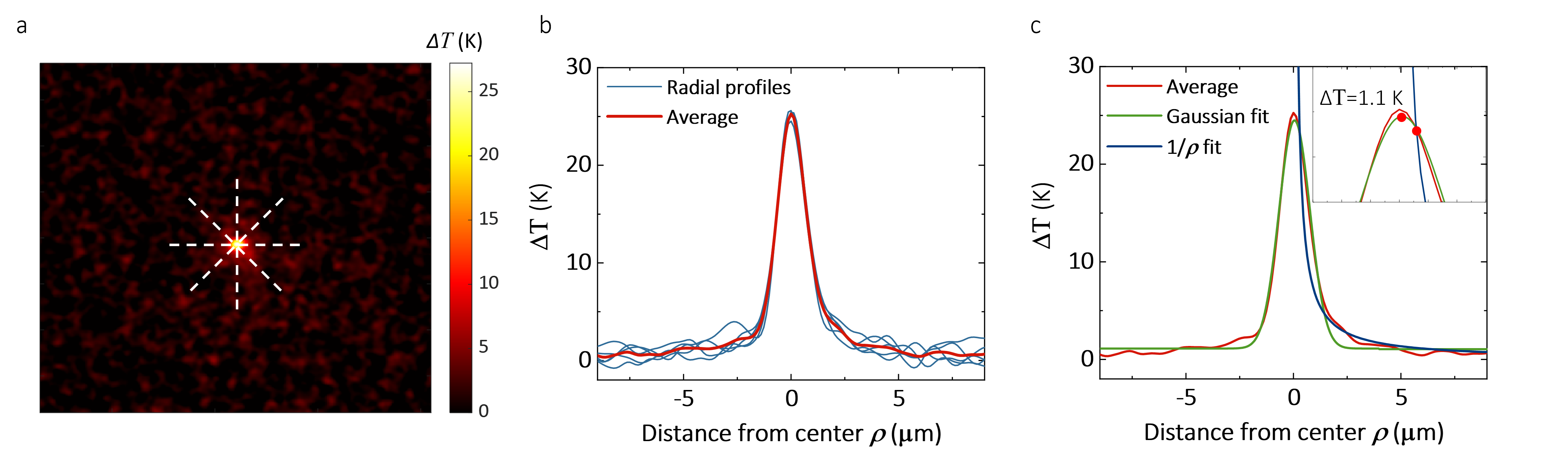}
\captionof{figure}{Temperature maps analysis overview. (a) Example temperature map of a single particle, with four line profiles centered on it. (b) Individual temperature profiles extracted from each dashed line in Figure (a) are plotted in blue, with the superimposed averaged profile in red. (c) Averaged temperature profile from Figure (b) plotted together with the relative Gaussian and $\sim$1/$\rho$ fit. Red data points show the resulting absolute temperature increase from the two different approaches, and the negligible deviation between them.}
\label{S4}
\end{figure}

\section{Quantification of average GNPs facet length}

\begin{figure} [h!]
\centering
\includegraphics[width=1\linewidth]{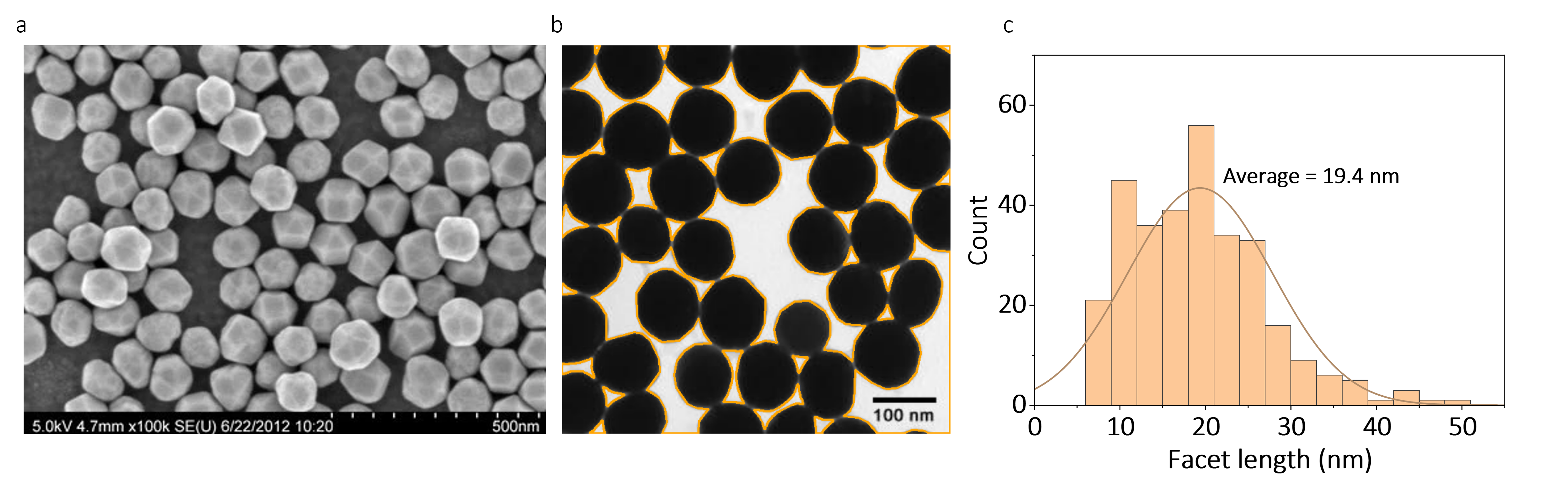}
\captionof{figure}{Example EM images of 100 nm diameter gold nanoparticles from Cytodiagnostics \cite{Cytodiagnostics} (a) and nanoComposix \cite{nanoComposix} (b) In the image, the outline of the detected particles is overlayed in orange. (c) Histogram of the extracted facets' length and the resulting average.}
\label{S5}
\end{figure}

To validate our assumption on the faceted shape of the employed gold nanoparticles, reference data are taken from electron microscopy (EM) data from two gold nanoparticle producers, namely nanoComposix \cite{nanoComposix} (Figure \ref{S5}a) and Cytodiagnostics \cite{Cytodiagnostics} (Figure \ref{S5}b). The images are post-processed with a Python script to identify the contour of the particles and fit it with a polygonal line. The extracted facets are plotted in the histogram in Figure \ref{S5}c, resulting in an average facet length of 19.4 nm. 

\newpage

\section{Mathematical model for hBN thickness characterization}

We employ the same mathematical model introduced by Khadir et al. \cite{Khadir2017SI} to retrieve the thickness of hBN layers from OPD measurements. The 2D material is considered as a slab with physical thickness $d$ and complex refractive index $\underline{n}=n+i\kappa$, where $n$ is the refractive index and $\kappa$ the extinction coefficient. The thin layer is at the interface between two media with respective refractive indexes $n_1$ and $n_2$. The expression of the complex transmission coefficient, resulting from the multiple reflections of the incident light within the slab, can be written as

$$
\underline{t}=\frac{4 n_1 \underline{n} e^{i k_0 d \underline{n}}}{(\underline{n}+n_1)(\underline{n}+n_2)-(\underline{n}-n_1)(\underline{n}-n_2)e^{i 2 k_0 d \underline{n}}},
$$

where $k_0=\frac{2\pi}{\lambda}$ is the wavevector of the light in vacuum. By normalizing this quantity by the transmission coefficient in the absence of the 2D material ($\underline{n}=n_1$), which reads $\underline{t}_0=\frac{2n_1}{n_1+n_2}e^{ik_0dn_1}$, we get

$$
\frac{\underline{t}}{\underline{t}_0}=\frac{2 \underline{n} (n_1+n_2)e^{i k_0 d(\underline{n}-n_1)}}{(\underline{n}+n_1)(\underline{n}+n_2)-(\underline{n}-n_1)(\underline{n}-n_2)e^{i 2 k_0 d \underline{n}}}.
$$

The complex quantity $\underline{t}/\underline{t}_0$ can be determined by QLSI from the measurements of the transmittance $T$ and the wavefront distortion $\delta l$, following

$$
\frac{\underline{t}}{\underline{t}_0}=\sqrt{T} e^{i\frac{2\pi}{\lambda}\delta l}
$$

and the hBN thickness $d$ obtained by fitting the complex equation.

\section{Annealing of GNPs}

To investigate the effect of the contact area on the resulting cooling factor, measurements were performed on the same set of hBN flakes before and after annealing. The sample was placed in an oven and heated up to $200\ \ ^\circ C$. After 40 minutes, the heating was stopped, and once the temperature was reached, the sample was taken out and measured. Our interpretation of the results, shown below, suggests a reshaping of the gold nanosphere upon annealing, resulting in a lower contact area with the underneath heatsink and a resulting decrease of the cooling factor. Considering the stability of the hBN layer upon temperature increase, up to $450\ ^\circ C$ in the cleaning procedure after the flakes transfer, in fact, rules out other possible mechanisms involved in the cooling factor difference.

\begin{figure} [h!]
\centering
\includegraphics[width=0.8\linewidth]{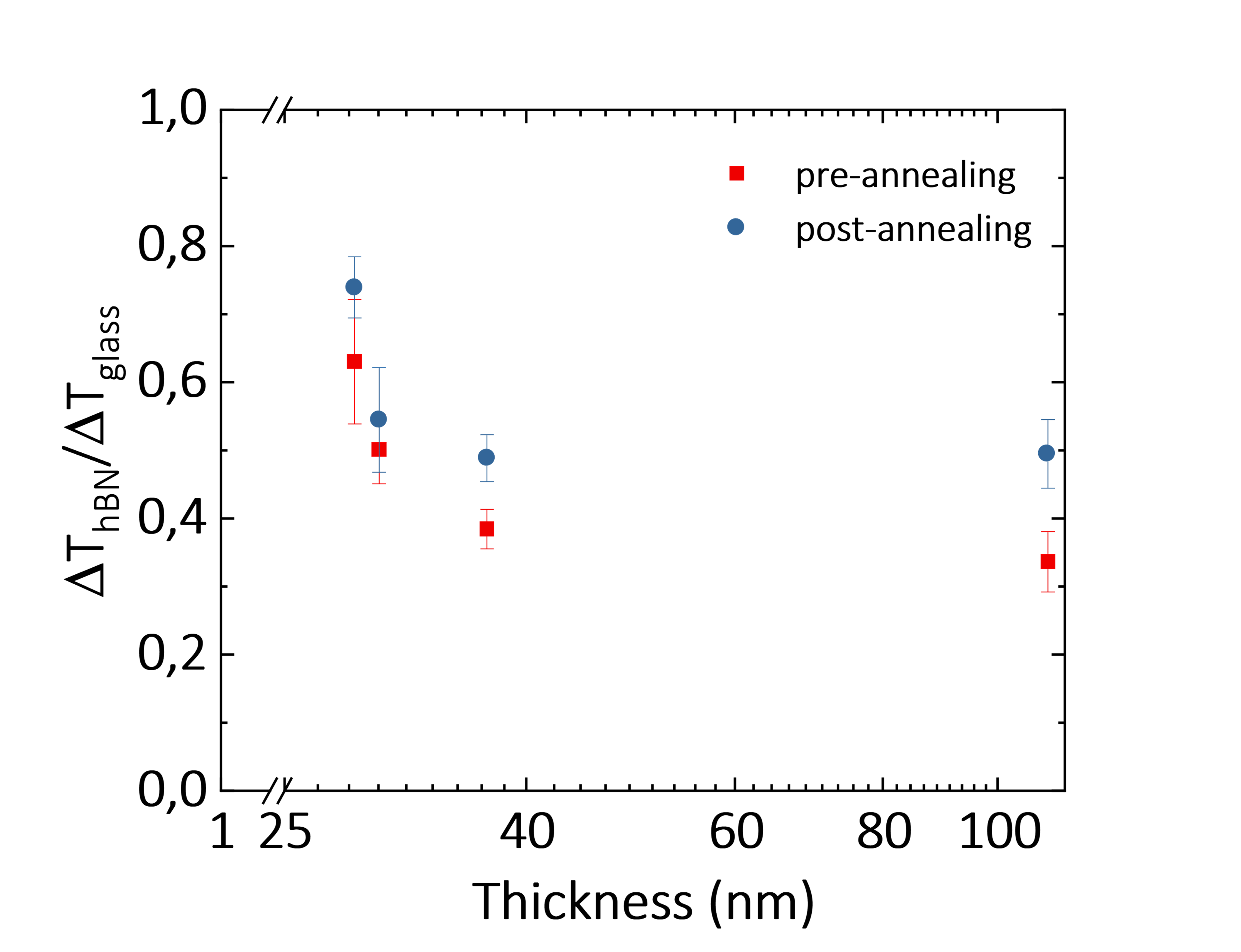}
\captionof{figure}{Experimental cooling factor for a set of hBN thicknesses before and after annealing of the gold nanospheres.}
\label{S6}
\end{figure}

\newpage

\bibliographystyle{unsrt}
\bibliography{BibliographySI}